# Progress towards quantitative design principles of multicellular systems

Eduardo P. Olimpio[1,2], Diego R. Gomez-Alvarez [1,2], and Hyun Youk[1,2,3]

[1]Department of Bionanoscience,
[2]Kavli Institute of Nanoscience, Delft University of Technology, Delft 2628CJ, the Netherlands
[3]Correspondence: Email: h.youk@tudelft.nl

**ABSTRACT**

Living systems, particularly multicellular systems, often seem hopelessly complex. But recent studies have suggested that beneath this complexity, there may be unifying quantitative principles that we are only now starting to unravel. All cells interact with their environments and with other cells. Communication among cells is a primary means for cells to interact with each other. The complexity of these multicellular systems, due to the large numbers of cells and the diversity of intracellular and intercellular interactions, makes understanding multicellular systems a daunting task. To overcome this challenge, we will likely need judicious simplifications and conceptual frameworks that can reveal design principles that are shared among diverse multicellular systems. Here we review some recent progress towards developing such frameworks.

**Concise definition of subject**

One of the important challenges in biology is quantitatively explaining how multicellular systems' behaviours arise from the genetic circuits inside each cell and the interactions among these cells. Communication among cells is one of the primary means for cells to interact with each other. One cell can influence how the other cell behaves by "talking" to that cell, for example, through a secretion of a signalling molecule that the other cell can respond to. Each cell can typically communicate with multiple cells located at various locations. Thus we can represent multicellular systems as complex communication grids. Discovering common principles that govern such multicellular communication grids is crucial for tying together a wide range of multicellular systems such as tissues, embryos and populations of microbes, under a common quantitative framework. But it has been difficult to find such principles. One difficulty is that we do not yet have generally applicable strategies for judiciously reducing the number of parameters in a multicellular system with a large number of intracellular and intercellular components. Another difficulty is that we usually do not know what quantitative metrics can characterize multicellular behaviors of interest. For example, it is unclear in many multicellular systems what quantities one should use to measure or model the degree to which cells in a tissue coordinate their behaviours to regulate the expression level of a gene that is common to them. Here we outline some possible methods for overcoming these difficulties and review recent studies that suggest that these methods may be potent.

**1. Towards quantitative design principles of multicellular systems**

That "more is different" is particularly evident in living systems. Multicellular systems, composed of many cells that interact with each other, involve a complex web of intercellular interactions that are layered on top of myriads of intracellular interactions [1–3]. For many natural systems, using first principles to derive how collective behaviors of cells arise through a hierarchy of interactions - going from genetic circuits to multicellular behaviors - is an outstanding challenge [4–6]. In this review, we outline these challenges, describe several studies that tackled these problems, particularly by using nascent approaches of bottom-up synthetic biology and systems biology, and based on these previous works we also suggest potential strategies for resolving several of the presented challenges. Taken together, the current studies suggest that the bottom-up approach, in which one starts from natural or engineered genetic circuits and then connect their behaviors to the behaviors of cell populations that those circuits control, is a promising way to understand how multicellular behaviors arise from the underlying molecular and cellular interactions [7, 8]. Here we will describe studies in which, with fairly simple experiments, researchers have mapped the relationships between intracellular circuits and the multicellular behaviors that they enable [9–13]. Moreover, mathematical modelling has been proven to be extremely valuable for a complete understanding of intracellular networks and their connections to the intercellular environment [14–17]. Supported by the recent progress, we posit that efforts should be made in devising new theoretical frameworks that can bridge the gap between individual cells and multicellular behaviors as well as finding common quantitative biological principles that are linked at different spatial scales (e.g., from intracellular to whole cell populations) and temporal scales (e.g., fast intracellular dynamics with slower changes that occur during mammalian development).

**2. Breaking multicellular systems into distinct functional and spatial modules may be possible**

Multicellular systems can range in many sizes. Perhaps one of the most interesting aspects about multicellular systems, one that fascinates researchers and suggests a strategy for deconstructing multicellular systems into simpler elements, is that a multicellular system can be composed of several other sub-systems, each of which can also be multicellular [18, 19]. For example, consider the human brain (Figure 1A). Many neurons and other types of cells (e.g., glial cells) make up the human brain. In a simplified picture, we can consider the brain to be composed of networks of neurons. On the other hand, we can also view the brain as consisting of many sub-modules (i.e., different regions of the brain that have different functions) such as the hypothalamus, cortex, amygdala and so on, each of which execute distinct functions and are connected to each other [20]. Such decomposition of a multicellular system into different modules can be in terms of the functions that each module carries out or in terms of the spatial locations of the groups of cells (e.g., cells are together in a given region of space). The two methods of decompositions do not always lead to the same modules. For instance, one can choose to divide the brain hemispheres into spatially (and arbitrarily) defined lobes (e.g. frontal, central, parietal, etc. [21]), or in terms of functions that are associated with those groups of cells (e.g. memory region, speech region, etc.). One can think of a computer as an analogy. A computer could be seen as a single, integrated electronic network formed by an intricate and almost endless arrangement of resistors, transistors and other electronic components. But often it is more useful to think of the computer as a collection of different sub-systems that are linked together. These sub-systems may be defined in terms of their function (e.g. screen: displaying data, keyboard: data input), or in terms of their spatial location (e.g. all the circuits on the motherboard). As another example, consider a tree (Figure 1B). It is a macroscopic multicellular system, consisting of many parts. For the sake of argument, we can say that it consists of leaves at the top of a tree trunk and roots at the base of the trunk. Here again, we can decompose the tree into distinct modules defined by their function or spatial locations.

In principle, the choice of how to sub-divide a certain system is arbitrary. However, some choices are more convenient than others in the sense that the different regions or divisions might have a certain degree of independence from each other. This is what we mean by *modularity* [22–26]. By construction, the electronic circuits inside a computer are highly modular. It is possible to replace the motherboard, the mouse or a USB port without altering the functioning of the system. Even inside a single microchip, the electric components are carefully put in place to minimize mutual interference, which results in a very high degree of independence. Crucially, the computer is designed to be modular by humans in a bottom-up manner.

In living systems, however, the situation is more complicated. One main reason is that we do not know, *a priori*, how a cell has been designed in a bottom-up fashion. The interior of a cell is so densely packed that interactions among its myriad components are nearly unavoidable. The presence of a transcription factor intended to regulate the expression of a certain gene might, for instance, block the transcription of a neighboring gene just by sitting close to it in the chromosome. Due to such lack of real or perceived independence among intracellular components, it is often unclear how one can deconstruct different intracellular pathways into different functional modules and if doing so is even possible. However, as we leave the world of individual cells and move to cell circuits that consist of networks of interacting cells, it may be possible that the spatial locations of cells dictate whether certain cells can be grouped into a single module or not. Of course, identifying these modules is not a trivial task. Unlike in the cartoons of the human brain or the tree (Figures 1A and 1B) there are no boundaries or dotted lines that divide different parts of multicellular systems. Nevertheless, we already know of some examples of modular multicellular systems, as has been proved by, for instance, successful cases of organ transplants to people. This suggests that the occurrence of modularity in multicellular systems might not be such an unusual phenomenon after all [24, 26, 27].

These simple ideas suggest that in order to understand the behavior of a macroscopic multicellular system, we could first try to find and understand each of its individual modules and then link the modules together to explain the behavior of the whole system. Multicellular behaviors among microbes are currently more amenable to systems-level approaches because there are fewer layers of interactions to explore and genetic engineering of microbes to perturb intra- and inter-cellular interactions is easier than doing the same for higher organisms [15, 28–30]. An illustrative system with few layers of interaction is that of "collective" bacterial resistance (Figure 1C). Here a single bacterium that is resistant to an antibiotic can induce antibiotic resistance of its neighbors by secreting a molecule (indole) that diffuses to the other cells to confer them resistance to the antibiotic [31]. Another classical example of a switch from an independent to a multicellular behavior is found in soil amoebae *Dictyostelium discoideum* that aggregate into a fruiting body when they are starved of nutrients (Figure 1D) [9]. In a broader picture, we can view these systems as model systems for researchers who wish to explore how multicellular behaviors could have evolved and what principles may underlie them. These examples all involve cells that live at the boundary of being unicellular and multicellular. In the case of *D. discoideum,* the cells live as individuals until they are starved of nutrients. In the case of the indole secreting bacteria, each cell maintains their individuality until a few cells begin to secrete indole.

Even for relatively "simple" microbes such as these, the boundaries between being unicellular and multicellular are often difficult to deduce because the interactions among the microbes are shaped by the spatial location of each cell in complex ways. This is particularly evident in the case of the complex dynamics that governs the *D. discoideum* cells forming fruiting bodies [9]. Despite the complicated nature of the transition from unicellular to multicellular forms, these types of microbial systems all have regimes in which it is clear that the cells are living as a unicellular entity and regimes in which it is clear that the cells are living as a collective multicellular entity.

Unlike the previous examples, many multicellular systems such as tissues and organs do not exhibit a transition from unicellular to multicellular forms. By definition, tissues and organs are composed of multiple cells that are supposed to work together. But a recent work suggests that it would be a mistake to simply assume that cells form a collective unit just because they are together in a tissue or an organ [14]. Consider, for example, the islets of Langerhans in the pancreas. The way the four major cell types ($\alpha$–cells, $\beta$–cells, $\gamma$–cells, and $\delta$–cells) are distributed over space is different in the islets of Langerhans of mice compared to that of humans (Figure 1E). The cells of different types at different locations communicate with each other through signaling molecules such as insulin. Each cell type secretes a different signaling molecule. It is thought that the different spatial distributions of the cells in the islets of Langerhans may affect the pancreas' function. But exactly how altering this "communication grid" (e.g., by rearranging the different cell types over space) would affect the pancreatic function is unknown because of the complexity of this multicellular grid. As in the case of the microbial systems mentioned above, one way to understand the coupling between cell-cell communication and spatial arrangements of distinct cell types in the islets of Langerhans may be to find modular structures in the islets of Langerhans. Recent work shows that cells can be completely autonomous or maintain some partial degree of autonomy even though they are nearly touching each other and exchanging signaling molecules in multicellular structures such as the islets of Langerhans [14]. The recent study by Maire and Youk [14] provides a conceptual and a practical framework for quantifying the amount of autonomy and the amount of collectiveness of cells. Using such a quantitative framework, we may be able to decompose systems like the islets of Langerhans into subsystems as in the case of *D. discoideum* or the indole-secreting bacterial system.

## 3. Communication among cells as a means of cell-cell interaction

Multicellularity requires coordination among cells. Cells primarily coordinate their behaviors by communicating with each other. To do so, a wide variety of cells, from bacteria to mammalian cells, secrete and sense each others' signaling molecules [32–35]. Despite the diversity of cell types, signaling molecules, and downstream functions inside the cells controlled by a variety of extracellular signaling molecules, we can broadly classify cell-cell communication into three types: autocrine, paracrine and juxtacrine (Figure 2). Autocrine signaling involves just one type of a cell. It occurs when a cell secretes a signaling molecule and at the same time produces a receptor that detects this molecule. This form of signaling not only allows the cell to engage in two-way communication with other cells that have the same receptor (which we call "neighbor-communication"), it also enables a form of "self-communication", in which the cell "talks" to itself by capturing the molecule that it has just secreted (Figure 2A) [10, 36]. For example, $CD4^+$ T-cells communicate with each other by secreting and sensing several molecules such as IL-2 that promote both the proliferation and death (apoptosis) of the cells (Figure 2B) [11]. A recent study has reported [10] that autocrine-signaling cells may exhibit different levels of "sociability", ranging from being completely asocial (only self-communication occurs) to being completely social (only neighbor-communication occurs). Researchers have shown through experiments on engineered yeast cells and mathematical models that cells tune where they are on this sociability spectrum by varying factors such as cell population density, secretion rate of signaling molecules and expression level of the receptor [10]. By understanding how cells can tune their social behavior, it is possible to link intracellular components to the multicellular behavior, and gain insight on how this transition occurs.

The soil amoebae *Dictyostelium discoideum* displays one of the most well-known examples of a transition from being asocial to being social (Figure 1D). When nutrients are abundant, the amoebae behave as asocial individuals. However, upon starvation they gather at one location in space and aggregate into a fruiting body. The fruiting body, like a tower with an observation deck, consists of a stalk and a nearly spherical dome at the top (Figure 1D). The cells that form the dome could be blown by wind to a nutrient rich region far away. But the cells that form the stalk cannot be blown away by wind because they are firmly planted on the ground in order to provide a structural support for the cells of the dome. Their sacrifice for the survival of the cells in the dome provides a case study for cellular altruism. The process by which the amoebae aggregate is guided by the signaling molecule cyclic Adenosine Mono-Phosphate (cAMP), which the amoebae simultaneously secrete and sense [9]. These cells are genetically programmed to produce more cAMP when they sense it. Thus in this case the factor that controls the cells' transition from asocial to social is the cell population density. In a nutshell, every time the cells come closer to each other, the concentration of cAMP in that region increases, and hence the signal that drives them to gather is amplified. The details of *D. discodeum*'s gathering process are actually more complicated than this [9], and we will discuss them further in section 6. The phenomenon of a cell population responding to changes in its density is so important and widespread that it has been given a name of its own, *quorum sensing* [35–37]. The aggregation of *D. discodeum* is a canonical example of a special type of quorum sensing called "dynamical quorum sensing" [38].

Paracrine signaling occurs when there are at least two types of cells, one that secretes a signaling molecule and another cell type that detects this molecule. Thus paracrine signaling is a form of "neighbor-communication" that we introduced earlier, in which one type of cell "speaks" and another type of cell "listens" (Figure 2C). Typically, this is a one-way communication because the listening cell does not talk back to the speaking cell. This mechanism is used, for example, by cancer cells to recruit $Cb11b^+Gr1^+$ myeloid cells to the tumor site, which helps to promote tumor growth [39]. Both the autocrine signaling and the paracrine signaling involve the diffusion of a signaling molecule and there can be disagreements and confusions on where to draw the line between the two [36]. In particular, there has been confusion on what autocrine signaling really is. Historically and strictly speaking, the distinction between the autocrine and the paracrine signaling relies on which cell secretes a signaling molecule and which cell makes a receptor that detects the signaling molecule. In this textbook definition, a cell that secretes a signaling molecule as well as makes a receptor for the molecule engages in autocrine signaling. But note that this definition does not specify who talks to whom. In a population of genetically identical autocrine cells, one cell can talk to another cell (by sending a signaling molecule to the other cell) as well as talk to itself (by capturing the signaling molecule that it just secreted outside moments ago). It is important not to confuse the molecular parts of a cell with the type of communication it engages in (self- and neighbor-communication) [10, 14, 36]. Paracrine signaling involves at least two types of cells, a sender and a receiver, where the sender secretes the signaling molecule but cannot sense it and the receiver only has the receptors but does not secrete the signaling molecule. Thus paracrine signaling means a neighbor-communication.

The third major form of cellular communication called "juxtacrine signaling" involves two cells communicating with each other through a direct physical contact between them (Figure 2E). A canonical example of juxtacrine signaling is the Notch-Delta signaling pathway [40, 41]. A careful examination of juxtacrine signaling revealed that juxtacrine signaling can involve both self- and neighbor-communication, just as in autocrine signaling. For example, in the case of Notch-Delta signaling, a cell produces both the Notch receptor and the Delta ligand, both of which sit at the membrane of the cell. Delta and Notch are designed in such a way that a cell's Delta ligand can bind to a neighboring cell's Notch receptor, which can trigger the expression of certain genes in the nucleus of the neighboring cell. In other words, a cell can use its Delta ligand to send signals to a specific neighboring cell that it's in direct physical contact with. Recent synthetic engineering of the Notch-Delta communication system has quantitatively investigated how a cell can inhibit itself from talking to its neighbors by plugging its own Notch receptors with its own Delta ligands [42]. In their engineered cell lines, the researchers discovered that this self-inhibition prevents the cell from talking and enable it to only listen to its nearest neighbors. By means of this self-inhibition mechanism that is built into the Notch-Delta pathway, some organisms, such as the fruit fly *Drosophila melanogaster,* are capable of generating precise spatial patterns during their development (Figure 2F) [43–45]. Therefore, a cell can tune its social behavior by the production of Delta molecules that will block its Notch pathway. The design principle underlying this tunable social behavior is similar to that of autocrine signaling, in which a cell "plugs" its own receptors by secreting signaling molecules and then immediately capturing those molecules with its receptors. This creates a tunable "background signal" that must be overcome by the neighboring cells if they want to be "listened to" [10, 36]. In other words, for such a self-inhibiting autocrine cell to listen to its neighboring cells, the neighboring cells must increase their secretion rate of the signaling molecule so that the concentration created by the neighboring cells is sufficiently higher than the concentration of the molecule created by the cell itself [10, 36].

Some biological systems do not fit directly into autocrine, paracrine, and juxtacrine signaling. For example, neurons communicate by electrical signals instead of using diffusing peptides, proteins, or by means of proteins at the junctions of direct physical contacts. But even in many of these examples, it is likely that the principles that govern their cell-cell communication is not very different from those that govern the self- and neighbor-communications in autocrine, paracrine, and juxtacrine signaling. Moreover the fact that we can classify a diverse zoo of cell-cell communication into the aforementioned three classes, and that there are intimate connections even among these three classes as illustrated above, suggest that there may be common principles that govern diverse cell-cell communications that we may be able to quantitatively describe. However, finding such all-encompassing mathematical or even qualitative descriptions still remains a challenging task.

## 4. Making sense of the combinatorial possibilities due to many ways that cells can be arranged in space

A multicellular system can form a vast number of possible networks due to the way that the cells are arranged in time and space and how the different cells communicate with each other. A useful first step is to simplify our analysis by making some assumptions. For instance, if we assume that cells can interact with each other by having the cells' locations to be fixed in some region of space, then potentially the total number of cell-cell interactions could be of a manageable size. Whereas if the cells are allowed to interact with everyone else in a population, as in the case of uniformly mixing liquid culture of microbes that are typical of laboratory culture settings, then we may think that there would be many more interactions than in the case of cells fixed in space. But this argument ignores the number of cell types that exist in a population.. The actual number of different types of interactions depends on both the spatial location of cells and the total number of types of interactions that are present in the population [46, 47]. The fact is that complexity is inherent even for the so-called "simple" multicellular systems and fully accounting for all possible cell-cell connections in an understandable model in which complex computer simulations are not required, is likely not possible. Therefore, the key to understanding design principles of multicellular systems lies in finding strategies for judiciously simplifying the typically large number of cell-cell interactions.

To make our argument more concrete, let us consider two types of cells, "cell A" and "cell B"(Figure 3A). They form a network consisting of two nodes and directed edges that connect them. A node can also be connected to itself. This means that the cell talks to itself as in the case of autocrine signaling. Each edge can be an arrow with either a blunt-end or a pointy end (Figure 3A). A pointy arrow represents a positive interaction. For example, a pointy arrow from A to B represents cell type A talking to cell type B to increase (activate) some intracellular event in cell type B, such as an expression of a particular gene. Conversely, a blunt-ended arrow from A to B represent cell type A talking to cell type B to decrease (repress) some intracellular event in cell type B such as an expression of a particular gene. "Node A" connecting itself with either type of arrows represents the cell regulating its own internal state, such as its gene expression by self-communication, as in autocrine signaling or Notch-Delta type juxtacrine signaling. There are $3^4 = 81$ possible types of networks even for this system consisting of just two cells because there are four possible edges, and each edge can be positive, negative, or non-existent (i.e., no interaction between the nodes). In fact, the actual number of networks is larger because there are many different parameters that characterize each edge [10, 16]. These parameters include the number of receptors present on the cell that receives the extracellular signal, the secretion rate of signaling molecules, and the intracellular processes that control the intracellular states after receiving the signaling molecule. We can think of this network as representing a multicellular system in which the spatial location of cells is unimportant. For example, this can represent a scenario in which the cells are moving around and interacting with everyone. It can also represent a case in which the cells are in a uniformly mixing liquid culture as is typical in laboratory settings in which bacteria and yeasts are grown in batch cultures. In such uniformly mixing liquid cultures, if the mixing is fast enough, all cells can communicate with each other and only the mean distance between cells matters.

If we now consider multicellular systems in which the cells are either fixed in their place (e.g., a colony of yeast cells) or are mobile in such a way that, unlike in the uniformly mixing liquid cultures, the cells do not randomly visit all possible locations in space (e.g., as in tissues), then the spatial location of cells becomes important. In such systems, even with only two types of cells, the possible number of spatial arrangements of the cells can be enormously large (Figure 3B). The number of possible configurations increases further if we now include the positive and negative interactions between cells that we previously mentioned (Figure 3A). This demonstrates how even small populations of cells with idealized multicellular interactions can lead to an enormous number of outcomes.

As a demonstration of a relatively small number of cells using the large number of connections among them to achieve numerous functions, we need to look no further than the worm *Caenorhabditis Elegans*. Every male worm is made of 1031 cells. Each cell performs different functions at different parts of the worm. With mere 1031 cells, the worm accomplishes many remarkable tasks, for instance, associative and non-associative learning [48]. Although we know a lot about the *C. Elegans'* genome and the molecules that are involved in the communication of cells within the *C. Elegans,* it is still unclear how the cells within the worm coordinate each others' actions and what quantitative principles underlie the coordination among the cells.

The *C. elegans* illustrates why we need a conceptual framework that lets us translate the known cell-cell connections into functional outcomes. Knowing all the connections among cells and the genetic circuits that govern those cells are insufficient for understanding how the group of cells function as a cohesive unit. But suppose we have a theoretical framework that takes the genetic circuits inside individual cells and the cell-cell connections (Figure 3) as inputs, and then as an output, gives us the total number of connections among the cells and the number of different cellular functions that they can achieve [14]. This means that we would potentially have a mathematical formula for computing the number of multicellular functions as well as how that number changes as we tune the spatial arrangements of cells and the different elements of the genetic circuits in those cells. Thus we would be able to link the molecular circuits inside cells with biological functions that the group of cells can achieve. Such a framework may be crucial for understanding how the mere 1031 cells that form the male *C. elegans* can achieve such a diversity of complex functions. For this reason, we think that going back to the basics in which we think of graphs that represent cell-cell connections (Figure 3) is crucial for obtaining design principles of multicellular systems.

**5. From individual cells to collective behaviors of cell populations**

Researchers have made much progress on revealing how multicellular organisms could have emerged through evolution [49, 50]. Moreover, applications of ideas from engineering and mathematical models have provided additional insights on how groups of cells function as cohesive multicellular entities [51–53]. Recent experimental and theoretical studies suggest promising bottom-up and top-down approaches for multicellular behaviors.

In order to understand multicellular behaviors, the main challenge we face is showing how population-level phenotypes arise from the phenotypes of individual cells [54–56]. Measurements on individual cells to learn how they behave is often easier [57, 58] than deducing the correct quantitative metric for describing the how the population as a whole behaves. This is currently a major challenge because we do not yet have a general strategy for bridging the gaps between the intracellular pathways inside cells with the behaviors of a group of cells. Succeeding in bridging, step-by-step, these two disparate realms would help us see how complexity emerges in a population of cells, step-by-step, taking individual cells as a starting point. As an example, a single resistant bacterium can induce resistance in part of the population by secreting indole [31] (Figure 1C). This example shows that having the full information about how a single cell develops its resistance is not enough to understand how the population as a whole develops resistance to an antibiotic. A step-by-step method for connecting the resistance of a single renegade cell to antibiotics with the resulting resistance of the whole population to the antibiotics would have important practical outcomes [59].

One population-level phenotype that has recently been studied is the homeostasis of a cell population density. Multicellular systems such as tissues and organs need to control their cell population density. A faulty regulation of the proliferation and death of the cells in a tissue or an organ can lead to harmful outcomes such as the organs dying and tumors developing. Often, the population should not be allowed to grow to a population density that rapidly depletes all the nutrients in the environment. Therefore, mechanisms for keeping the number of cells below the carrying capacity of the environment (i.e., the point at which the nutrient is completely depleted) are important. Researchers have recently developed a mathematical model to explore "cell circuits" (networks of communicating cells) that enable "paradoxical behaviors", in which the binding of a ligand to the cell's receptor promotes both the cell's proliferation and death [16]. The researchers found that if the ligand promoted the proliferation and death rates proportionally in the right amount, the population of cells achieved homeostasis that is robust to changes in the initial concentration of the cells [16]. Moreover, the researchers confirmed this prediction of their mathematical model in a series of experiments on $CD4^+$ T-cells. Namely, the researchers cultured the $CD4^+$ T-cells that secreted and sensed the ligand IL-2. The T-cells use IL-2 as an autocrine signal, thus achieving both self-communication and neighbor-communication [10]. They then observed that the rate of cell death due to apoptosis depends linearly on the concentration of IL-2 [11]. On the other hand, the binding of IL-2 to the receptors on $CD4^+$ T-cells is responsible for their progression into replicative phases of the cell cycle [62]. This proliferation rate of the cells depends on the concentration of the IL-2 in a sigmoidal fashion [11]. The apoptosis rate of the cells depends on the concentration of the IL-2 in a linear fashion. These two "antagonistic" (also called "paradoxical") effects yields two stable steady-state values of the cell population density (Figure 4A): (1) Zero (extinction state), or (2) fixed homeostasis level that is below the carrying capacity of the environment. Therefore, the population of cells either (1) goes extinct or (2) expands or contracts until it reaches the homeostatic value [11]. Thus simply by secreting and sensing one signaling molecule, cells can regulate what population-level to reach and maintain over time. These studies [11, 16] form an example of how mathematical modelling and a simplified view of cell signaling deduced that out of a myriad possible cell circuits that one can have, only a few of them yield a particular biological function.

More than studying the components of a population to understand an observed function, it is also important to explore which functions we can be build from a given set of components. In engineering terms, instead of understanding the circuit of a mounted electronic device, we want to make the electronic device ourselves starting with basic components such as transistors. Such a bottom-up approach is important for understanding how population-level phenotypes arise from individual cells and genetic circuits inside the cells. Researchers have recently used this bottom-up approach to construct population-level functions [12, 13] (Figure 4B). By building genetic circuits in budding yeasts and E. coli cells that could function as logic (Boolean) gates liked the ones found in computers, the researchers showed how the cells could communicate with each other to make collective decisions. For example, as an input, the researchers' yeast cells responded to a variety of extracellular molecules (the mating factors from *S. Cerevisiae* and *C. Albicans*, glucose, doxycycline, etc.). As an output, the yeast cells turned on the expression of a gene encoding a fluorescent protein or turned on the secretion of a signaling molecule in a digital manner (i.e., the cell was either maximally fluorescent or not fluorescent, the cell either secreted the signaling molecule at a maximal rate or it did not secrete the molecule). The authors then used a mathematical theory to calculate the number of possible population-level logic functions that a population of such communicating cells could realize. By using only a few engineered cells (typically two to five) and signaling molecules (typically two to four), the researchers showed that it was possible to implement most of the logic functions with a few extracellular inputs (typically two to three). These studies showed how one could engineer different kinds of genetic circuits into E. coli and yeasts, mix the resulting different cell types in one environment, and as a result, realize distinct logic gates at a population-level.

Further studies that find relationships between the genetic circuits inside individual cells and their population-level behaviors would help us gain a fundamental understanding of how complexity arises in multicellular systems [58, 63]. They may also help us find practical applications of engineered multicellular systems. Recent papers have demonstrated such a potential by synthesizing complex bio-compounds through the division of labor among different cells in a population [64] and by patterning physical materials with cells [65].

**6. Tuning multicellular behaviors**

Progress towards obtaining quantitative design principles of multicellular systems requires understanding how individual cells tune their behaviors so that the entire population can execute a desired biological function. Depending on this function, the cell benefits from acting alone while in other occasions being part of a population is beneficial [66, 67]. Therefore, it is essential to find connections between the dynamics inside cells and its effects on the extracellular environment in order to understand how the cells coordinate their behaviors and maintains this coordination over time and space. The formation of the fruiting bodies by the soil amoeba *Dictyostelium discoideum* is a good example of this. When starved for nutrients, the soil amoeba coordinate their motion and secretion of cAMP to aggregate into a fruiting body (Figure 1D) [9]. The fruiting body releases their spores into the air, giving the cells in those spores a chance to colonize more hospitable environments. The dynamics of the intracellular pathway that leads to the formation of the fruiting body are not yet completely known. But we know that cell-cell coordination arises due to oscillations in the concentration of the cytosolic (intracellular) cAMP over time. Specifically, the period of this oscillation and the maximal concentration of cAMP produced during the oscillation control the asocial (freely roaming) to social (aggregated) transition of the soil amoebae. Researchers have shown that only oscillations with a sufficiently high maximal concentration of intracellular cAMP trigger the *D. discoideum* to enter into the aggregation phase [9]. They showed that cells secrete the cAMP to communicate with each other and that a positive feedback mechanism, in which the cells secrete more cAMP as they sense more cAMP, causes the maximal concentration of the intracellular cAMP to build up. This type of dynamics, called "excitable dynamics", is often seen in neurons that communicate with each other through electrical spikes. Having realized this, researchers have recently showed that the FitzHugh-Nagumo (FHN) model, also used to model neurons voltage spike and fire, presents the required properties and has shown good agreement with the experimental data [68]. Importantly, this model suggested a possible intracellular network that generates the excitable dynamics in the concentration of the intracellular cAMP (Figure 5A). Furthermore, the FHN model predicted that adding an extracellular cAMP to a population in which the *D. discoideum* cells' intracellular cAMP concentration oscillates in a synchronized manner would remove the synchronization. This would result in each cell oscillating autonomously and not in synchrony with each other. Indeed the researchers verified this prediction experimentally. This case study illustrates the importance of finding mathematical insights from the observations, which not only fits the experimental data but also provides deep conceptual insights that underlie the complex multicellular system [17, 69].

Another recent study that discovered simple yet general principles that govern multicellular behaviors used a theoretical investigation of "secrete-and-sense cells" [14]. Secrete-and-sense cells secrete a signaling molecule and produce a receptor that detects this molecule. Thus they can realize both self-communication and neighbor-communication [10, 36]. In this study, the researchers considered a hexagonal spatial arrangement of the secrete-and-sense cells. They then showed that one could rigorously define and quantify the "amount of autonomy" and the "amount of collectiveness" of cells. The secrete-and-sense cells secreted one type of signaling molecule. As a function of the sensed concentration of the extracellular signaling molecule, the cells could either turn "ON" the expression of a gene and simultaneously increase the production of the signaling molecule or turn "OFF" the expression level of a gene while simultaneously decreasing the secretion rate of the signaling molecule. Such a positive-feedback mediated secrete-and-sense cells are ubiquitous in nature. They are found in T-cells, cancer cells, and the *D. discoideum* [9, 11, 70]. Using both an analytical theory and cellular automata simulations, the researchers recently showed that spatial patterns such as stripes and islands could be formed by the ON- and OFF-cells [14]. These spatial patterns emerged as the cells varied the different parameters of the genetic circuits (e.g., secretion rate of the signaling molecule). Importantly, the researchers showed that they could count the total number of possible spatial patterns that could emerge later in time without knowing what each of the hundreds to thousands of cells were initially doing (i.e., without knowing the ON/OFF state of each cell at the initial time). The researchers then defined the concept called the "entropy of population", which quantified the total number of the spatial patterns that could be stably maintained over time [14]. They derived an analytical formula for the entropy of population that may be applicable to any generic secrete-and-sense cell populations in nature. The entropy of population is a function of the amount of the signaling molecules secreted by the cells, the amount of receptors possessed by the cells, and the concentration of the signaling molecule required for the transition between the OFF and the ON states to occur. Depending on the "signaling length" (i.e., the radius of the diffusive cloud of the signaling molecules that the secreting cell creates around itself), the researchers observed that there was a change in the degree of cellular autonomy (Figure 5B). In the case where the signaling length is small, each cell is less affected by the neighbors and is thus more autonomous. Here the entropy of population is higher. However, increasing the signaling length makes the cells less autonomous and more collective (as measured by their quantitative metric for the amounts of autonomy and collectiveness). In this case, there is more cell-cell coordination in the population [10]. This will form regions in which the ON cells are clustered together and the OFF cells are clustered together (i.e., patches of islands of ON/OFF cells) that reduce the entropy of population. The idea of defining the entropy of population, the amount of collectiveness of cells, and the amount of cellular autonomy in a quantitative manner illustrates a possible general approach of representing autonomous and collective behaviors of cells in other multicellular systems. Furthermore, cells can tune their sociability in various ways, for instance, by changing the amount of receptor, by inducing degradation of the signaling molecule, by chemically affecting the diffusion constant of the molecule or by changing the time scale of the interaction [10, 71, 72]. Therefore, the entropy-approach [14] establishes a framework to quantitatively understand how different phenotypes can arise from the changes in the behavior of individual cells in a population. Such an approach is likely to be fundamental to deconstructing the complexity of multicellular systems.

## 7. A new framework for quantitatively understanding multicellular systems

We have so far discussed the main difficulties in obtaining a quantitative understanding of multicellular systems and some recent studies that suggest potential solutions. It is worthwhile to emphasize that multicellular systems are inherently complex due to the myriad connections among cells of different types and many intracellular components that are typically present in these systems. Thus it may be that no single principle can unite diverse multicellular systems. Any simplification will overlook some aspects of the multicellular system and may not capture the full richness of the biological system at hand. But it is still worthwhile to not give up and search for general principles that may capture and unify even some simplified versions of various multicellular systems.

This challenge is similar to the one faced by physicists in describing physical systems with many particles. Physicists discovered mostly in the 19th century that they could use macroscopic parameters such as pressure and temperature to describe the behaviors of physical systems composed of many particles, such as magnets and gases. For instance, when studying the thermodynamics of gases, it is not necessary to calculate the position and momentum of each particle in the box of gases since macroscopic parameters like volume, pressure and temperature are often enough to explain the main features of the system (Figure 6A). In fact, it would be impossible to measure the position and velocity of every one of the Avogadro number of gas particles in the box. Such a description of the gas is possible due to a well-established framework, namely statistical mechanics, which connects macroscopic parameters with the microscopic details of the system. Statistical physics allows us to quantitatively understand how macroscopic behaviors of a system of many particles results from the microscopic components and interactions among those particles [73].

Although statistical physics has not yet proven to be useful in understanding multicellular systems beyond some trivial applications, it motivates us to search for similar approaches in order to simplify the complexity of multicellular systems and networks [74, 75]. For gases, pressure, volume and temperature were concepts that came before scientists even knew that a gas was composed of a collection of particles. In fact, it was even before most scientists accepted the existence of atoms. Several experiments had already been done using parameters like temperature long before Boltzmann and others developed the main pillars of statistical mechanics. Interestingly, the case for cells and multicellular systems is the opposite. Much research has been done to characterize the molecular components such as proteins and the detailed functions of those components inside cells. Indeed it is now possible to describe many of the cellular functions thanks to the advances in genomics, genetics, and molecular biology [76–78]. Moreover, researchers have recently made much progress in understanding how cells communicate and what this implies for their behavior [35, 36, 79, 80]. Yet a lot of questions still remain. A complete picture of our understanding of multicellular systems requires that we start thinking about how our current and vast knowledge of the workings of individual cells can be tied together to understand multicellular behaviors [25]. This requires a search for quantities that describe the main features of multicellular systems and their components (Figure 6B). Widely used concepts like evolvability, modularity, phenotype map and stress resistance, to name a few, could potentially be explored in a quantitative fashion to find common principles among different multicellular systems [25–27, 81–83]. This could provide new conceptual insights and, potentially, a new "statistical biology" could emerge as a solid framework that bridges the gaps between individual cells and the multicellular systems that they form.

    The model systems and the recent conceptual developments that we discussed in this review may be starting points for developing a new framework for multicellular systems. The multicellular systems involving yeast, E. coli and *D. discoideum* that we have discussed here are some of the simplest multicellular systems. We believe that simple systems like these have the potential to inspire a new theoretical framework (Figure 6B) just as the ideal gas has done for thermodynamics and statistical mechanics (Figure 6A). This is because these simple multicellular systems have relatively few components and interactions that involve at most one or two signaling molecules. This simplicity facilitates our search for a general framework. Moreover, we can easily tune various parameters in these systems in a way that is prohibitively difficult to do for more complex systems such as embryos and tissues. Although a framework that is developed from these simpler systems may not be immediately applicable to more complex systems such as embryos, we would likely be able to develop it further in order to find design principles that are conserved through evolution and thus are applicable to more

complex multicellular systems. Using the analogy with statistical mechanics, when introducing interactions in an ideal gas, we obtain the van der Waals equation which, although resembling the equation of an ideal gas, introduces drastically new features to the system such as phase transition. Furthermore, statistical mechanics that originated from ideal gas has developed over time to tackle complex phenomena such as fluid flows and magnetism. Fluids and magnets look different from the ideal gas but we can use the same framework (statistical mechanics) to study them. Equivalently, a "statistical biology", after it is successfully developed from the simple multicellular systems, may be a framework that we can extend to understand complex systems such as developing embryos and tissues.

Reasoning along these lines may help scientists working on both sides of the spectrum (intracellular systems and multicellular systems) to find a common conceptual ground. The recognition that living systems have particular characters that differ from those of non-living systems [25] suggests that we should actively search for a radically new conceptual framework for understanding multicellular systems. Recent progress in systems and synthetic biology that we have reviewed here suggests possible routes.

## ACKNOWLEDGMENTS

We thank the members of the Youk Laboratory for helpful discussions and comments. H.Y. is supported by a NWO NanoFront Grant and the ERC Starting Grant (677972-MultiCellSysBio). We apologize to researchers whose relevant work we may have missed in citing.

**Figure captions:**

**Figure 1. Exploring the concept of cell circuits.** Multicellular systems consist of cells that interact at different spatial scales. **A.** The human brain can be thought of as a network of communicating neurons and other types of cells (e.g., glial cells) that coordinate their behaviours through cell-cell communication so that they can achieve various systems-level functions such as memory and cognition. **B.** Cells at different parts of a tree, such as the leaf cells, trunk cells, and root cells, each perform distinct functions. But these functions must be integrated together for the tree to function as a whole organism. **C.** A subset of cells in a bacterial population secretes indole, which is a molecule that stimulates the efflux of antibiotic molecules. This in turn confers the entire population a resistance to the antibiotic, including those cells in the population that do not secrete indole. Thus a few cells in a population can tune the fitness of the entire population. **D.** Soil-living amoebae *Dictyostelium discoideum*, when starved, can communicate with each other through the secretion of cyclic Adenosine Mono-Phosphate (cAMP) to transition from being unicellular to a multicellular fruiting body. This behavior allows the *D. discoideum* to serve as a model system to understand the emergence of collective cellular behaviors from initially autonomous cells. **E.** The pancreatic cells responsible for hormone production are localized in regions known as the islets of Langerhans. The islets contain different types of cells that communicate with each other by secreting and sensing certain molecules (hormones). The hormones regulate the glucose level in the blood. Different proportions and spatial arrangements of the cells inside the islets result in variations in glucose regulation.

**Figure 2. Classifying communication among cells. A.** Autocrine signalling occurs when the same cell secretes and senses the signalling molecule. **B.** $CD4^+$ T-cells secrete and sense IL-2, a molecule that increases both the proliferation and the death of cells, leading to a homeostasis of cell population density (i.e., the cells robustly maintain a fixed cell population density over time). **C.** In a paracrine signalling, one cell secretes a signaling molecule that is sensed by another type of cell that has a receptor to detect

this molecule. **D.** Cancer cells recruit CD11$^+$Gr$^+$ myeloid cells using paracrine signalling of the CXCL1 molecule. These myeloid cells enhance the viability of cancer cells by producing S100A8/9 factors. **E.** Juxtacrine signalling occurs due to physical contacts between cells. At the cell-cell contact junction, signaling molecules on the two cell surfaces meet and enable one cell to communicate with the other cell. **F.** Communication through the molecules Notch and Delta is an example of juxtacrine signaling. The binding of a Delta ligand to a Notch receptor causes juxtacrine signaling. The Notch-Delta signaling is responsible for spatial pattern formation during development of embryos, such as in the patterning of the veins in the wings of the fruit fly, *Drosophila melanogaster*.

**Figure 3. Cell-cell communication generates complexity in multicellular systems. A.** In a system of two cells, each cell can "talk" to itself or to the other cell. Sensing of a signaling molecule regulates the intracellular state (e.g., expression of a gene) in two different ways: positive regulation or negative regulation. A cell does not have to talk to or listen to anyone. Thus there are a total of $3^4 = 81$ possible ways to "wire" these two cells. **B.** A multicellular system in which seven cells, each of which can be in one of two states ("ON" or "OFF") and placed on the vertices or the center of a hexagon. There are total of $2^7 = 128$ possible arrangements of such cells.

**Figure 4. Top-down and bottom-up approaches for deducing the rules that govern multicellular systems. A.** A top-down approach. Autocrine signaling through IL-2 allows CD4$^+$ T-cells to control both the rates of their proliferation and death (apoptosis). Among all possible genetic circuits that control the secretion of, sensing of, and response to IL-2, only a few of them enable the population to balance the proliferation and death rates in just the right way to maintain the cell population density at a fixed level over time. Such homeostasis of cell population density requires that the proliferation rate of cells increases nonlinearly as a function of the sensed concentration of the extracellular IL-2 while the death rate of the cells increases linearly as a function

of the extracellular IL-2 concentration. This creates a bistable system at a population-level: The cell population density can be stably maintained in one of two values - zero or a higher value. **B.** A bottom-up approach. Engineering different genetic circuits inside budding yeast to create different "cell types" (one type for each genetic circuit). The genetic circuits tune the secretion of, sensing of, and response to different signaling molecules. In this way, communication among the different cell types in a population was tuned to control the collective behaviors of the population consisting of distinct cell types. The signalling molecules act as wire among the different cells. This enables the population to realize different logic gates akin to the logic gates found in computers.

**Figure 5. Conceptual frameworks that reveal quantitative principles underlying the transition between collective and autonomous behaviors of cells. A.** A top-down approach. Unveiling the response of single cells from the population-level dynamics. In the soil-living amoebae *Dictyostelium discoideum*, the collective behavior of the cells, in which the cells aggregate to form a fruiting body, emerges due to cells communicating with each other by secreting and sensing the signaling molecule cyclic Adenosine Mono-Phosphate (cAMP). Although several molecular mechanisms responsible for the formation of the fruiting bodies remain unknown, researchers observed that the concentration of the cAMP inside and outside the cells over time and space had similar features as the dynamics of the voltage spikes observed in neuron-to-neuron communications. Motivated by this, the researchers could use FitzHugh-Nagumo model, which is also used to model neuron-to-neuron communication, could quantitatively capture the concentrations of cAMP over time and space, and thus the dynamic process through which the initially autonomous *D. discoideum* cells aggregated into fruiting bodies. **B.** A bottom-up approach. Secrete-and-sense cells secrete a signaling molecule and produce a receptor that detects this molecule. Thus they can realize both self-communication and neighbor-communication. In this study, the researchers considered a hexagonal spatial arrangement of the secrete-and-sense cells. In doing so, they showed that we can rigorously quantify the "amount of autonomy" and the "amount of collectiveness" of secrete-and-sense cells. The

researchers defined the concept called the "entropy of population", which quantified the total number of the spatial patterns that could be stably maintained over time. They derived an analytical formula of the entropy of population that may be applicable to any generic secrete-and-sense cell populations in nature.

**Figure 6. Need for new conceptual frameworks to understand multicellular systems.** How can we describe macroscopic physical and biological systems that are composed of an immense number of microscopic interactions? Can we deduce a new "statistical mechanics" that applies to cells? **A.** Knowing the position and momentum of every gas particle in a box is impractical because there are typically many Avogadro numbers of particles in the box. Instead, statistical mechanics uses macroscopic parameters such as pressure, temperature and volume to describe the macroscopic state of the box of particles. Statistical mechanics is thus a framework that ties the microscopic and macroscopic variables together by identifying a few key parameters that summarize the myriad interactions among the microscopic particles in a system. **B.** Recent studies, including the ones described in Figure 5, are beginning address how multicellular systems composed of many communicating cells could also be described by a few parameters that capture the myriad interactions among multiple cells in a population. These and further studies may yield a new conceptual framework that can bridge the gap between individual cells and the whole population through a few essential parameters, akin to how statistical mechanics uses pressure and temperature to describe the box of gas particles.

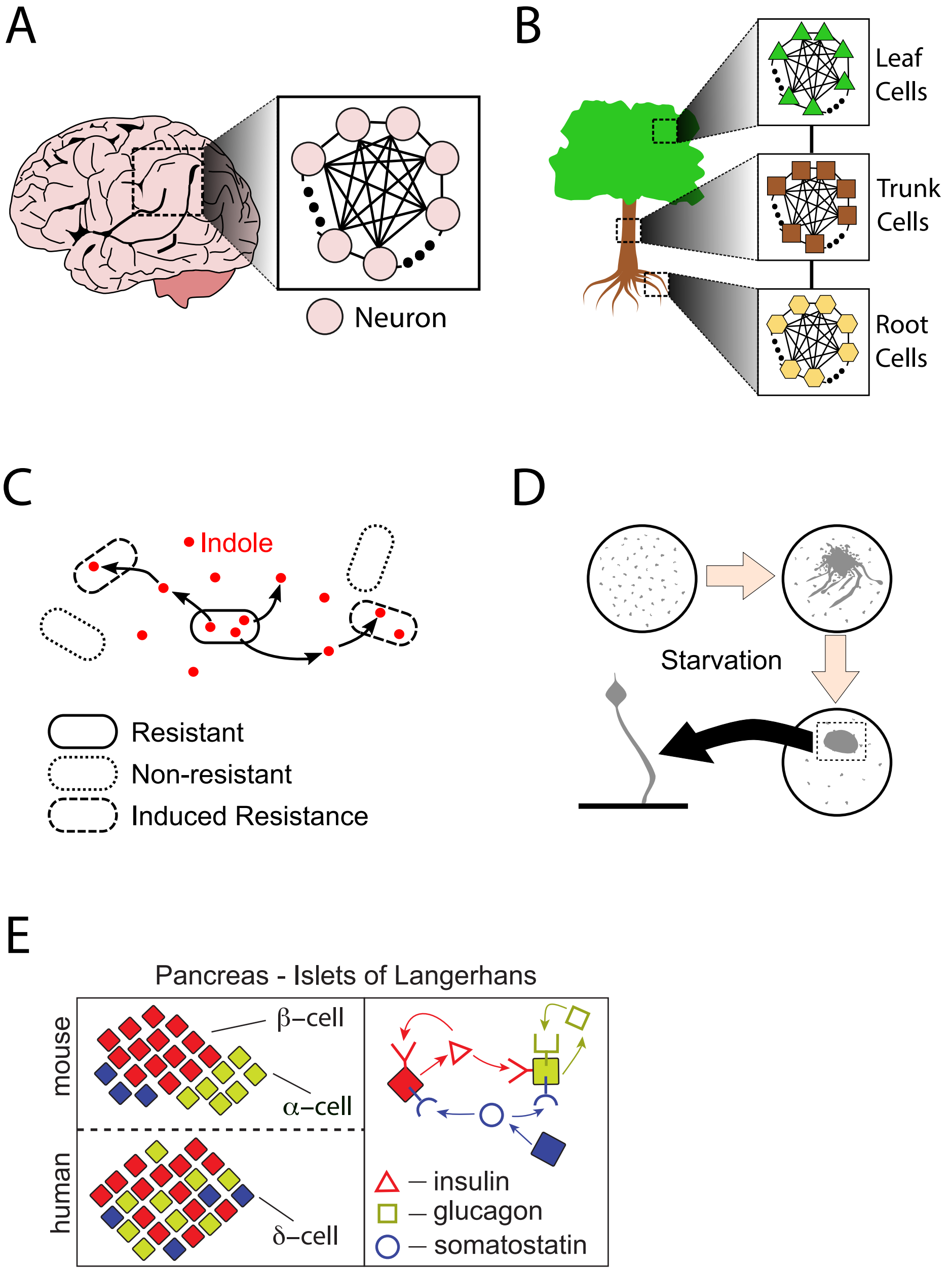

Fig. 1

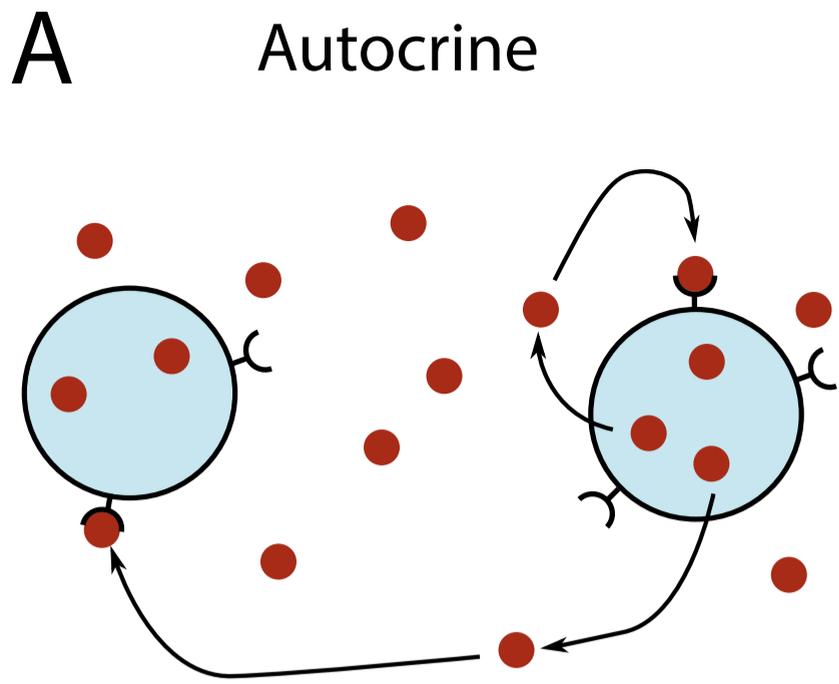
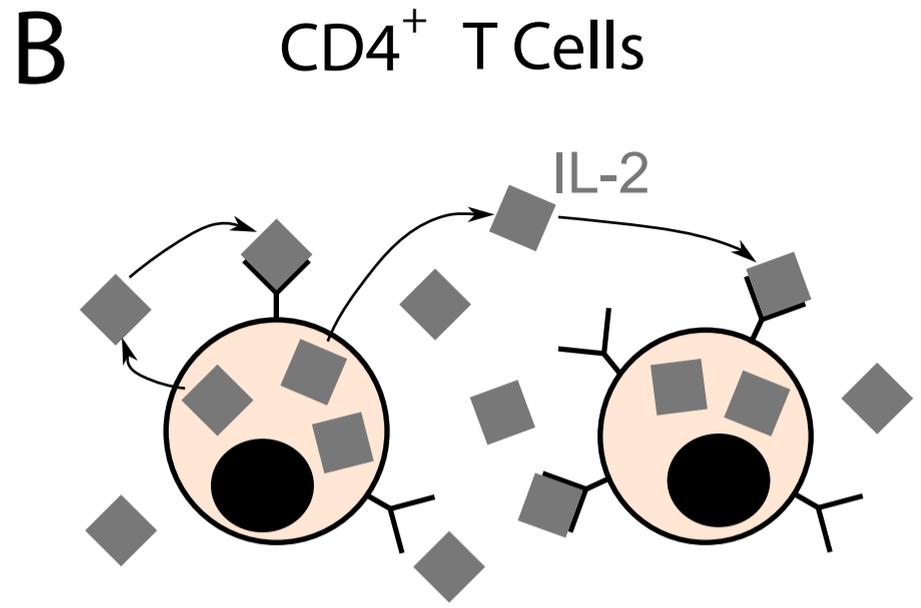
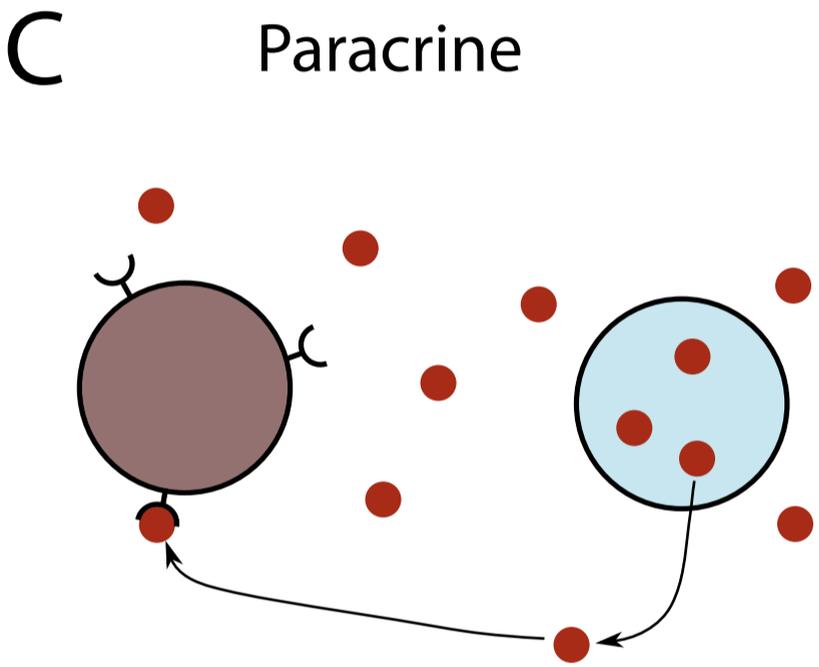
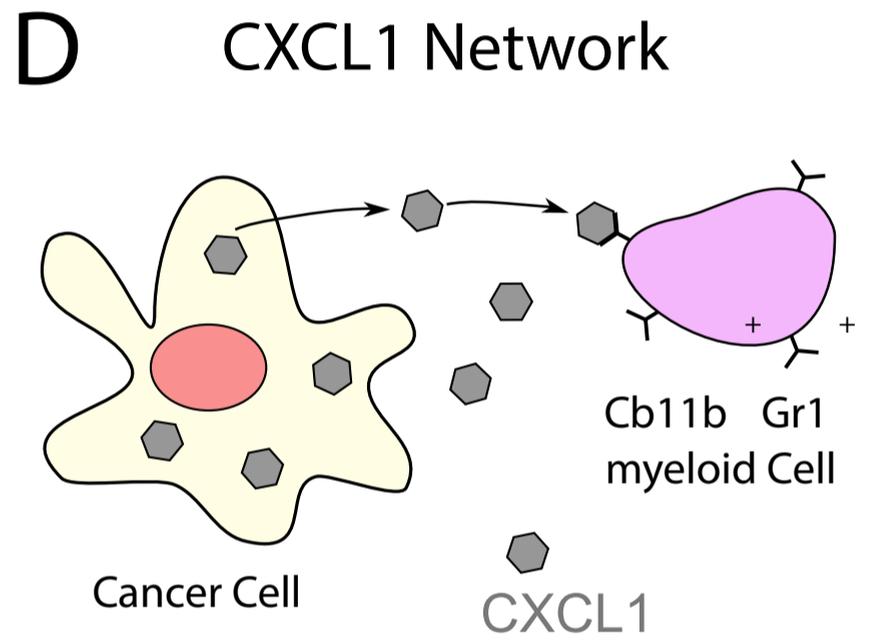
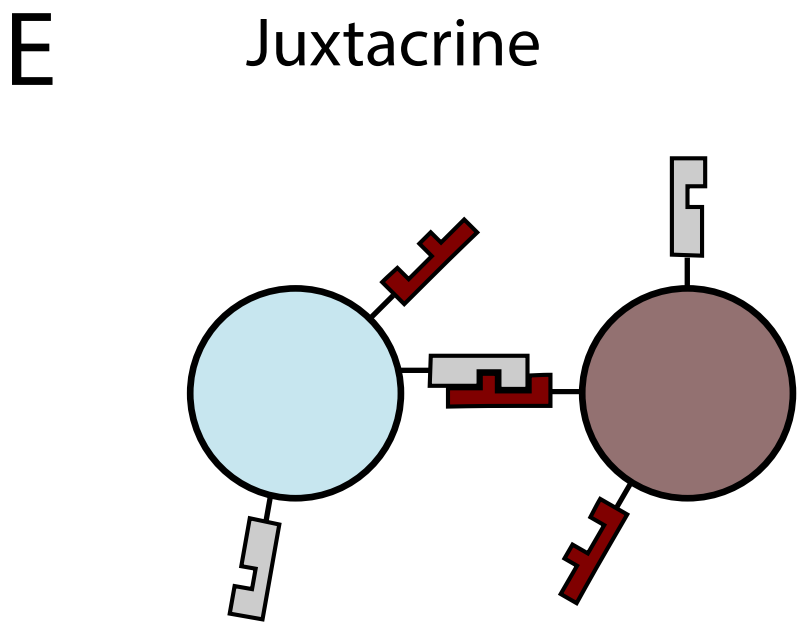
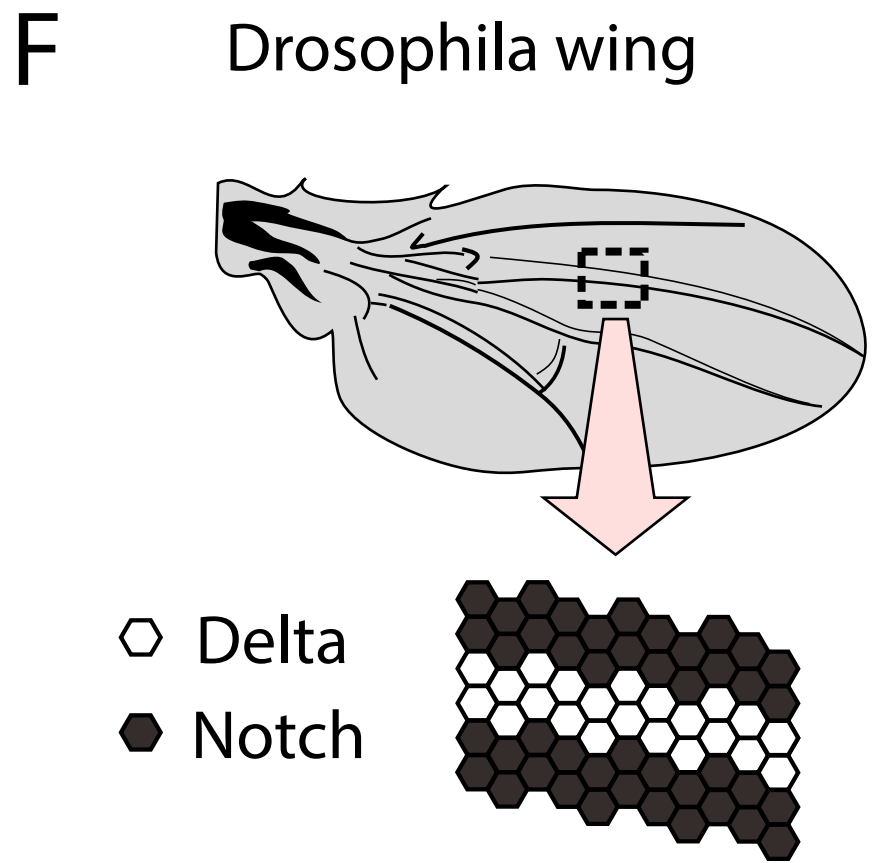

Fig. 2

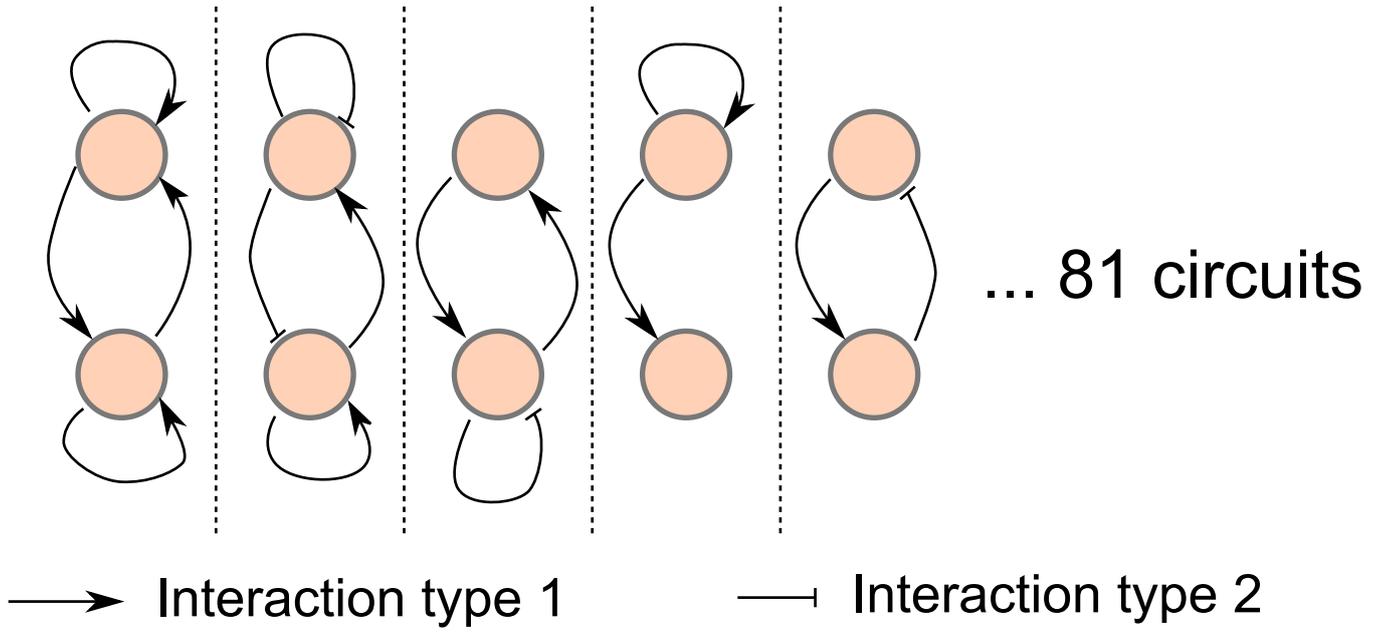

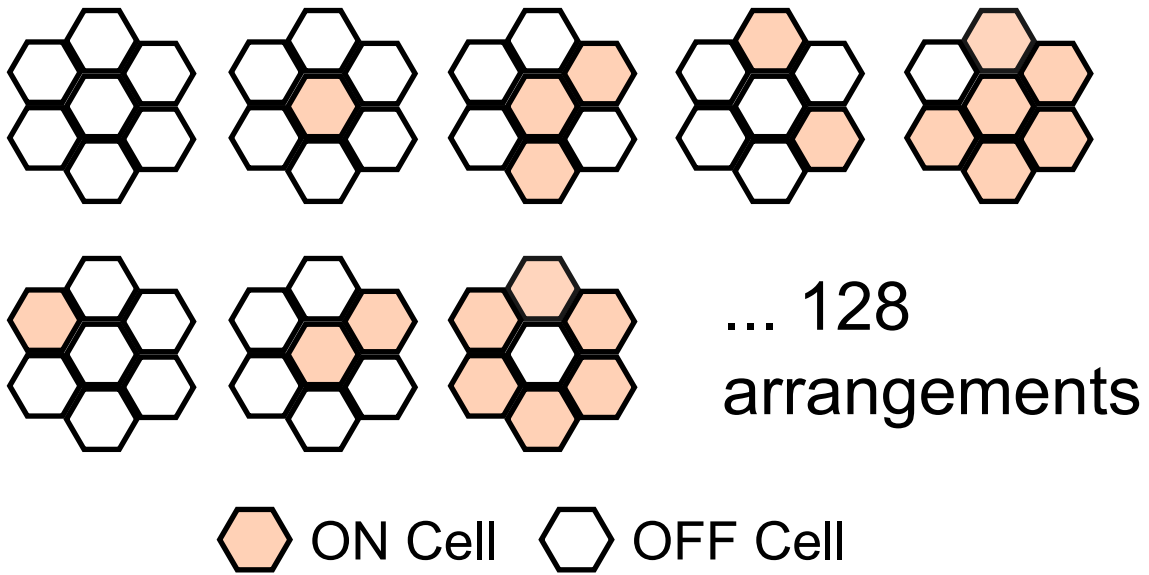

Fig. 3

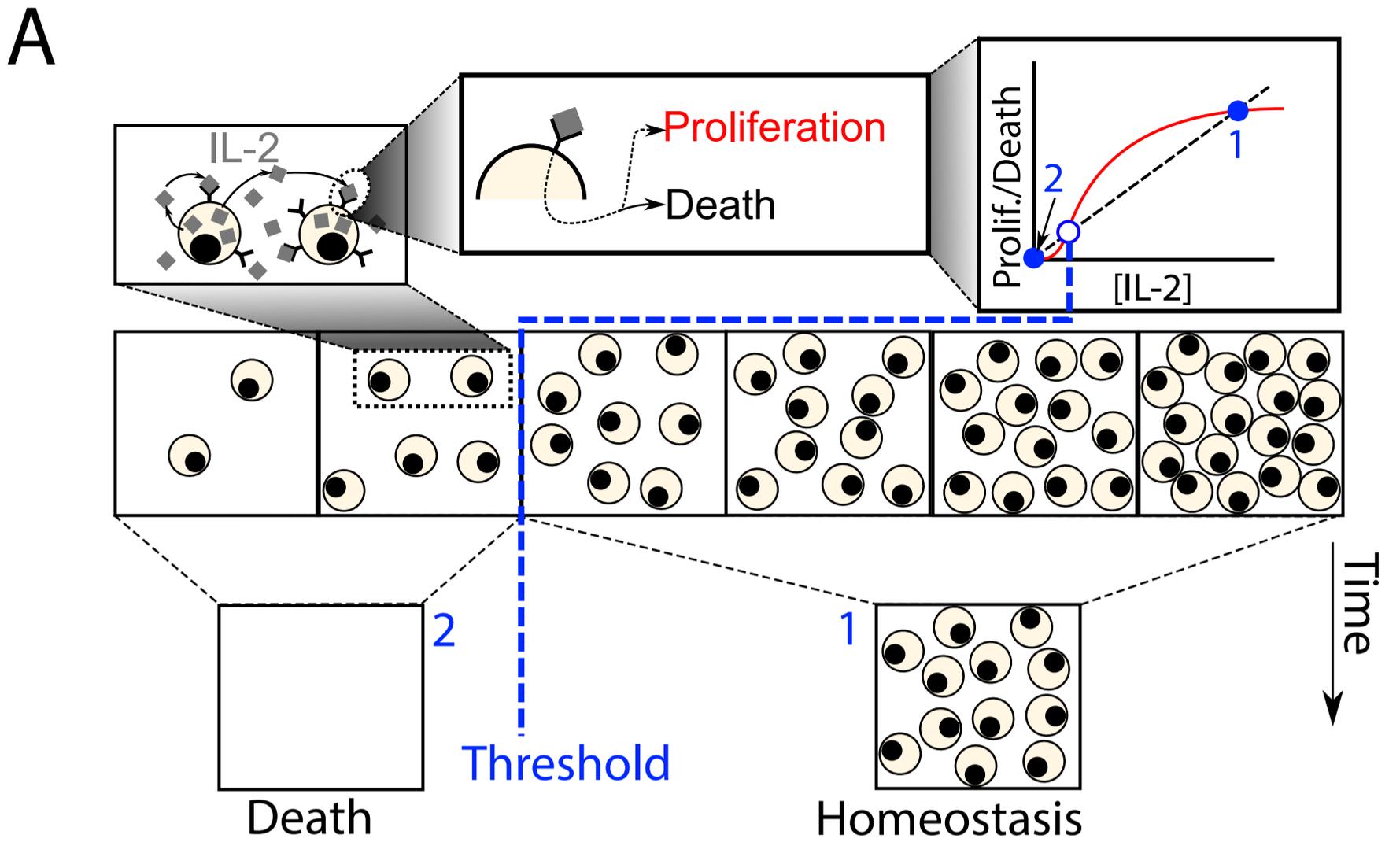
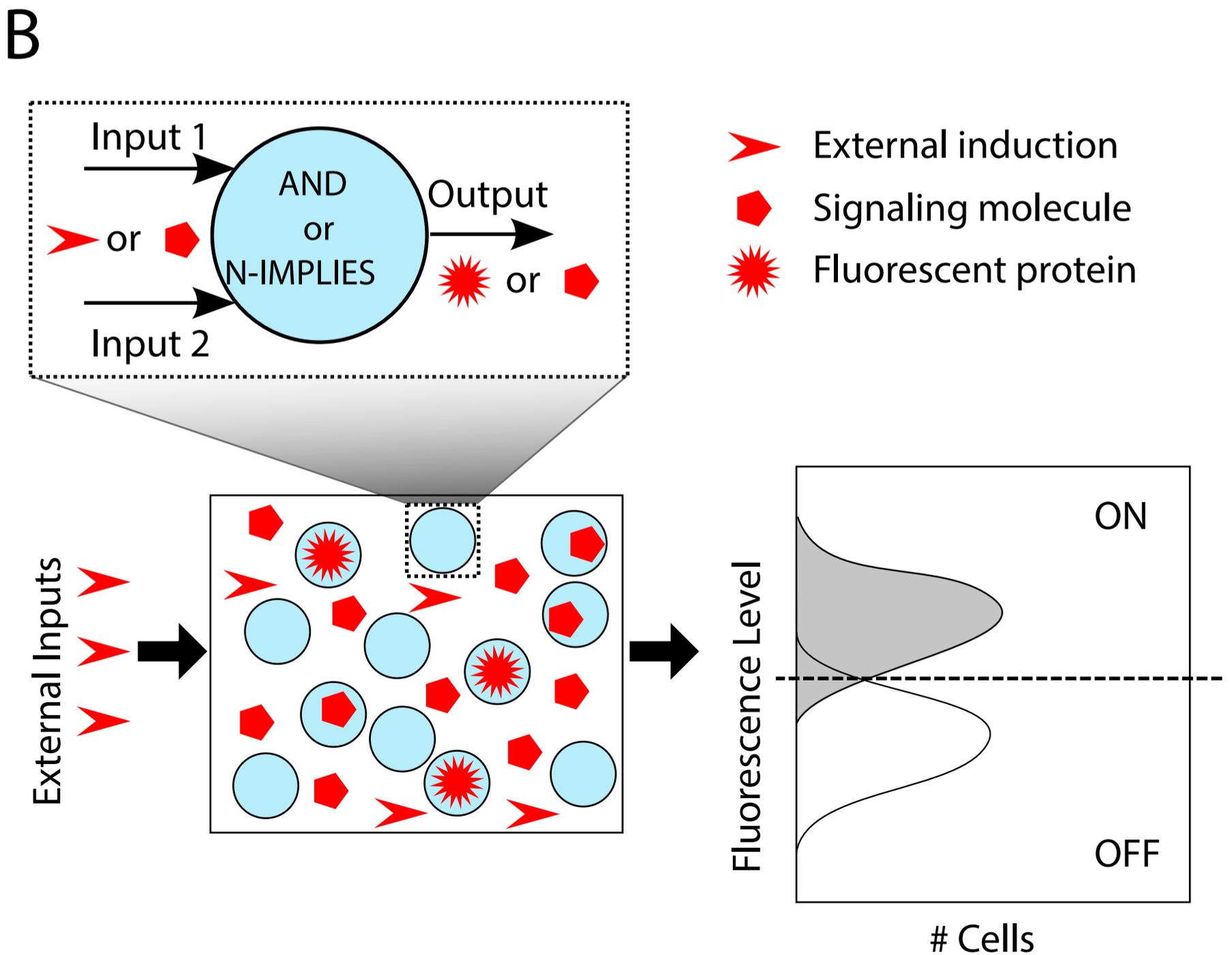

Fig. 4

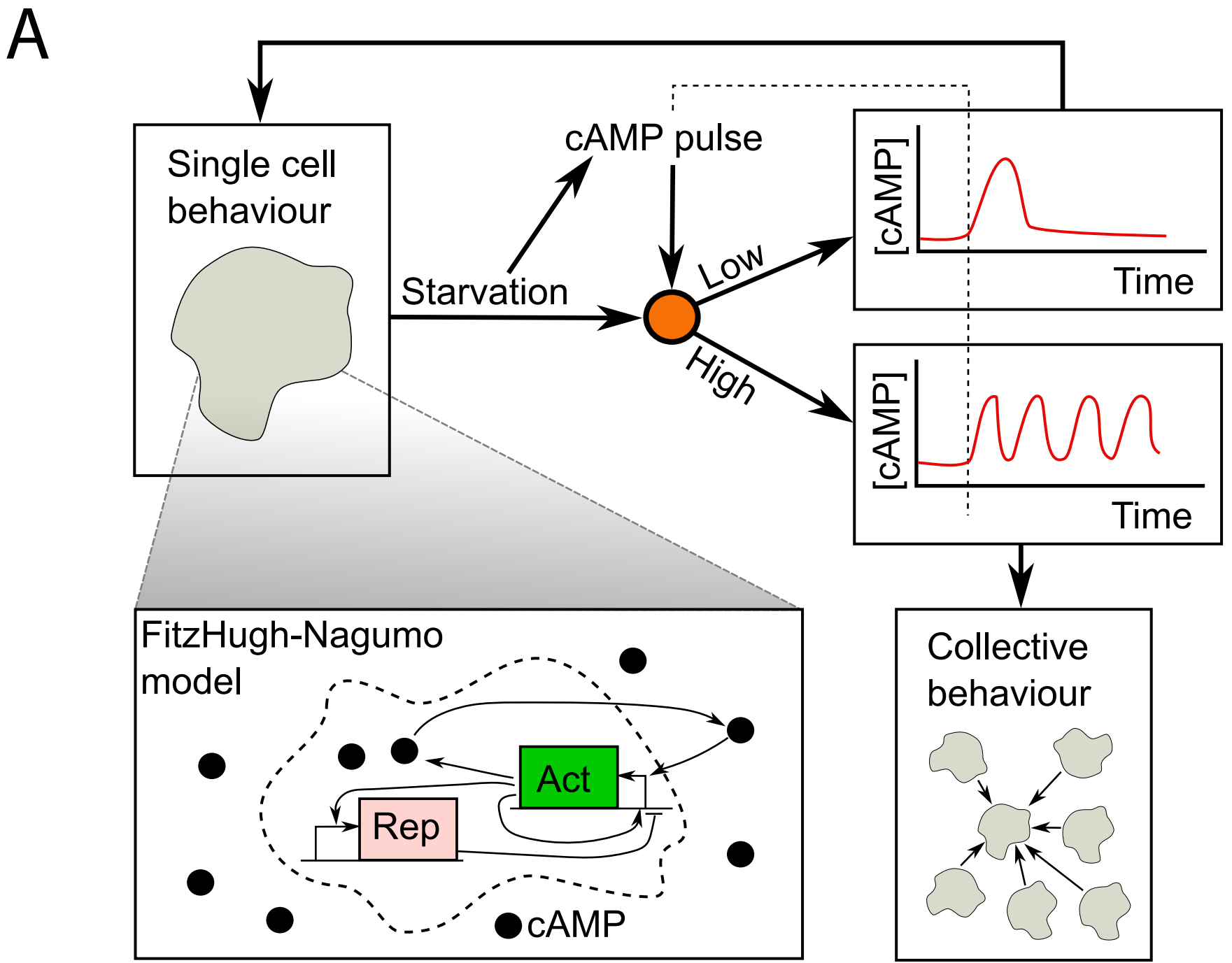
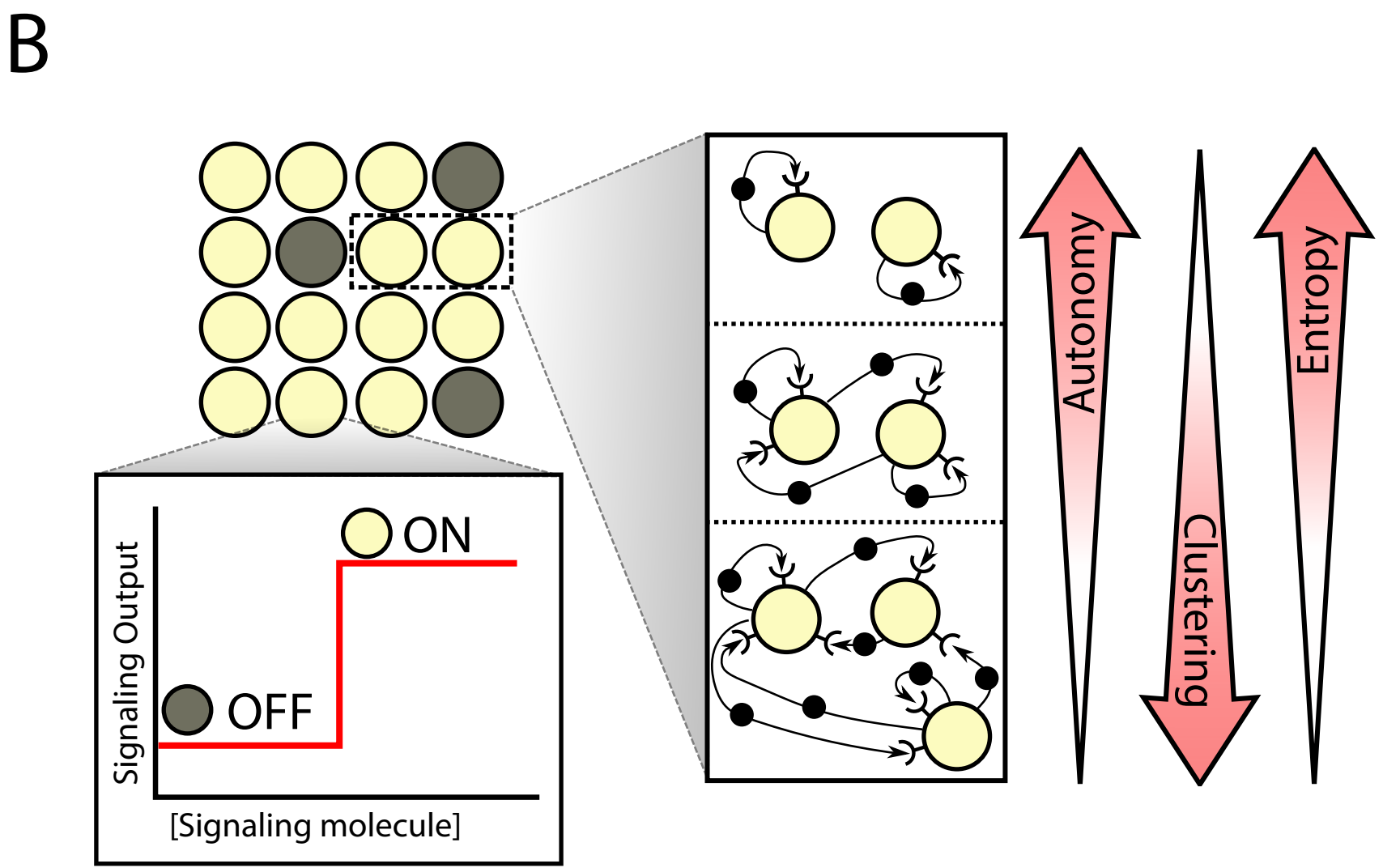

Fig. 5

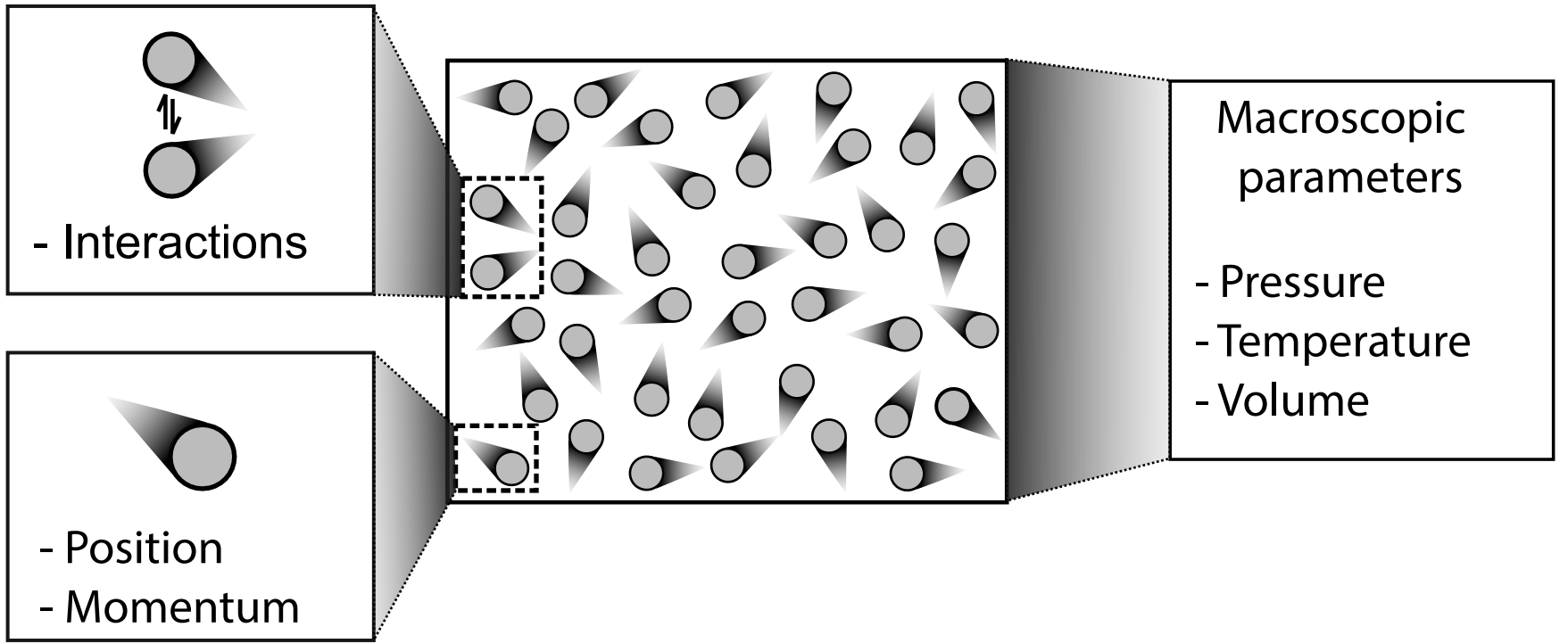
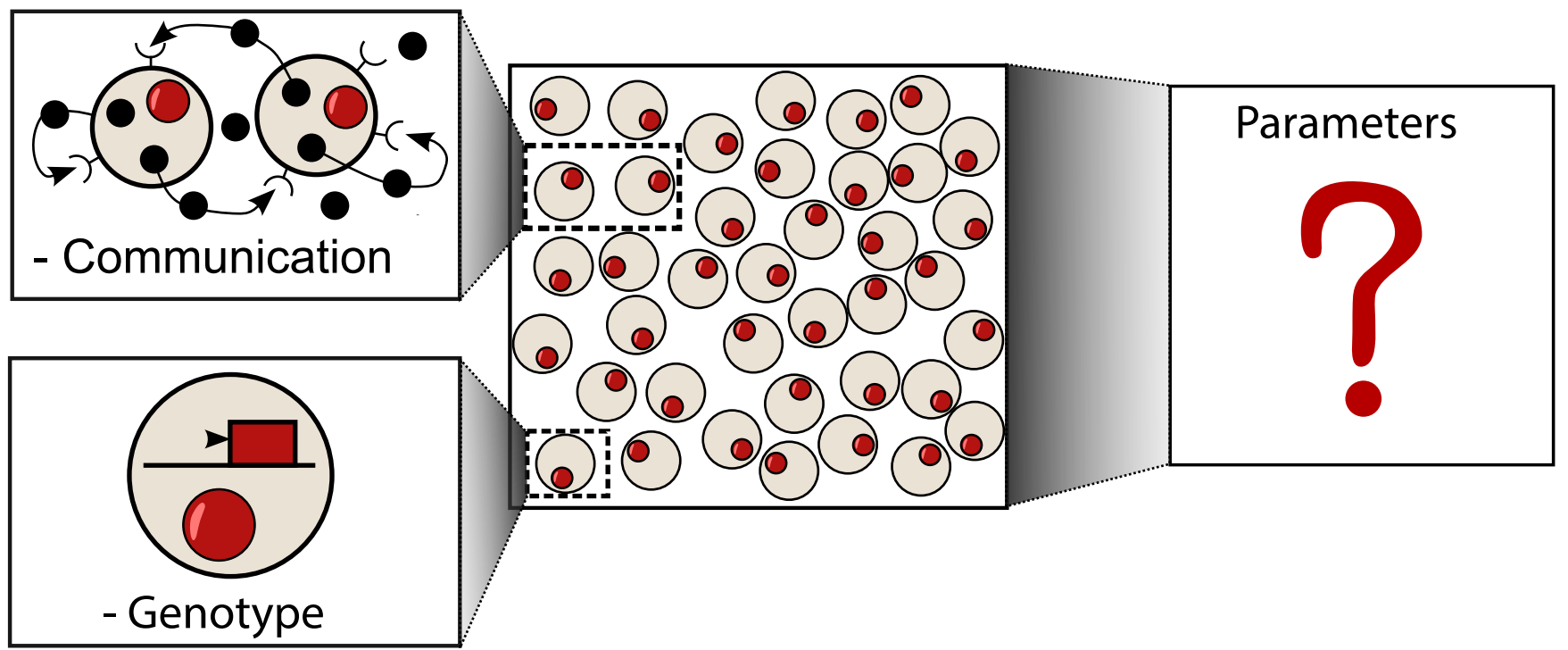

Fig. 6